\documentclass[11pt,a4paper]{article}

\usepackage[]{hyperref}
\usepackage{amsmath, amsthm, amssymb}
\usepackage{graphicx}

\def\expt(#1#2){\langle #2 \vert #1 \vert #2 \rangle}

\title{A puzzle for the field ontologists}

\author{Shan Gao
\\Research Center for Philosophy of Science and Technology, 
\\ Shanxi University, Taiyuan 030006, P. R. China
\\ E-mail:  \href{mailto:gaoshan2017@sxu.edu.cn}{gaoshan2017@sxu.edu.cn}.}

\begin{document}
\maketitle


\begin{abstract}\noindent 
It has been widely thought that the wave function describes a real, physical field in a realist interpretation of quantum mechanics. In this paper, I present a new analysis of the field ontology for the wave function. First, I argue that the non-existence of self-interactions for a quantum system such as an electron poses a puzzle for the field ontologists. If the wave function represents a physical field, then it seems odd that there are (electromagnetic and gravitational) interactions between the fields of two electrons but no interactions between two parts of the field of an electron. Next, I argue that the three solutions a field ontologist may provide are not fully satisfactory. Finally, I propose a solution of this puzzle that leads to a particle ontological interpretation of the wave function. 

\end{abstract}

\vspace{6mm}

\section{Introduction}

The field ontology for the wave function has been a popular position among philosophers of physics (Ney and Albert, 2013). 
For example, in Bohm's theory, the wave function may be regarded as either a real, physical field in a fundamental high-dimensional space (Bell, 1987, p.128; Albert, 1996, 2013, 2015), or a multi-field in three-dimensional space (Forrest, 1988, ch.5; Belot, 2012; Hubert and Romano, 2018). The former is usually called wave function realism. 
There are also other similar field ontologies of quantum mechanics. 
For example, in Everett's theory, spacetime state realism has been proposed (Wallace and Timpson, 2010; Wallace, 2012, ch.8; Swanson, 2018). According to this view, the fundamental ontology of quantum mechanics consists of a state-valued field evolving in four-dimensional spacetime. 
Besides, in collapse theories, the mass density ontology is a popular view (Ghirardi, Grassi and Benatti, 1995; Ghirardi, 1997, 2016; Allori et al, 2008, 2014). According to this view, ``what the theory is about, what is real `out there' at a given space point $x$, is just a field, i.e. a variable $m(x,t)$ given by the expectation value of the mass density operator $M(x)$ at $x$.'' (Ghirardi, 2016). 

There have been a few obections to the field ontologies of quantum mechanics, such as the objections to wave function realism (Monton, 2002, 2006, 2013; Lewis, 2004, 2013, 2016; Maudlin, 2013; Gao, 2017, ch.7, Chen, 2017), the objections to spacetime state realism (Arntzenius, 2012, ch.3; Baker, 2016; Ismael and Schaffer, 2016; Norsen, 2016), and the objections to the mass density ontology (Myvold, 2018).  
In this paper, I will present a new analysis of the field ontology for the wave function. 
First, I will argue that the non-existence of self-interactions for a quantum system such as an electron poses a puzzle for the field ontologists. 
If the wave function represents a physical field, then it seems odd that there are (electromagnetic and gravitational) interactions between the fields of two electrons but no interactions between two parts of the field of an electron.\footnote{I thank a referee of European Journal for Philosophy of Science for this precise expression of the puzzle.}  Next, I will argue that the three solutions a field ontologist may provide are not fully satisfactory. Finally, I will propose a solution of this puzzle which leads to a particle ontological interpretation of the wave function. 



\section{The puzzle}

Consider two electrons being in a product state at an initial instant. 
The wave function of electron $A$ is $\psi(x)={1\over {\sqrt{2}}}[\psi_1(x)+\psi_2(x)]$, where $\psi_1(x)$ and $\psi_2(x)$ are two normalized wave functions respectively localized in their ground states in two small identical boxes 1 and 2. 
The wave function of electron $B$ is $\psi(x)={1\over {\sqrt{2}}}[\varphi_1(x)+\varphi_2(x)]$, where $\varphi_1(x)$ and $\varphi_2(x)$ are two normalized wave functions respectively localized in their ground states in two small boxes 3 and 4 which are identical to the boxes 1 and 2. 

\begin{figure}[h]\label{4}
\begin{center} 
\includegraphics[scale=0.25]{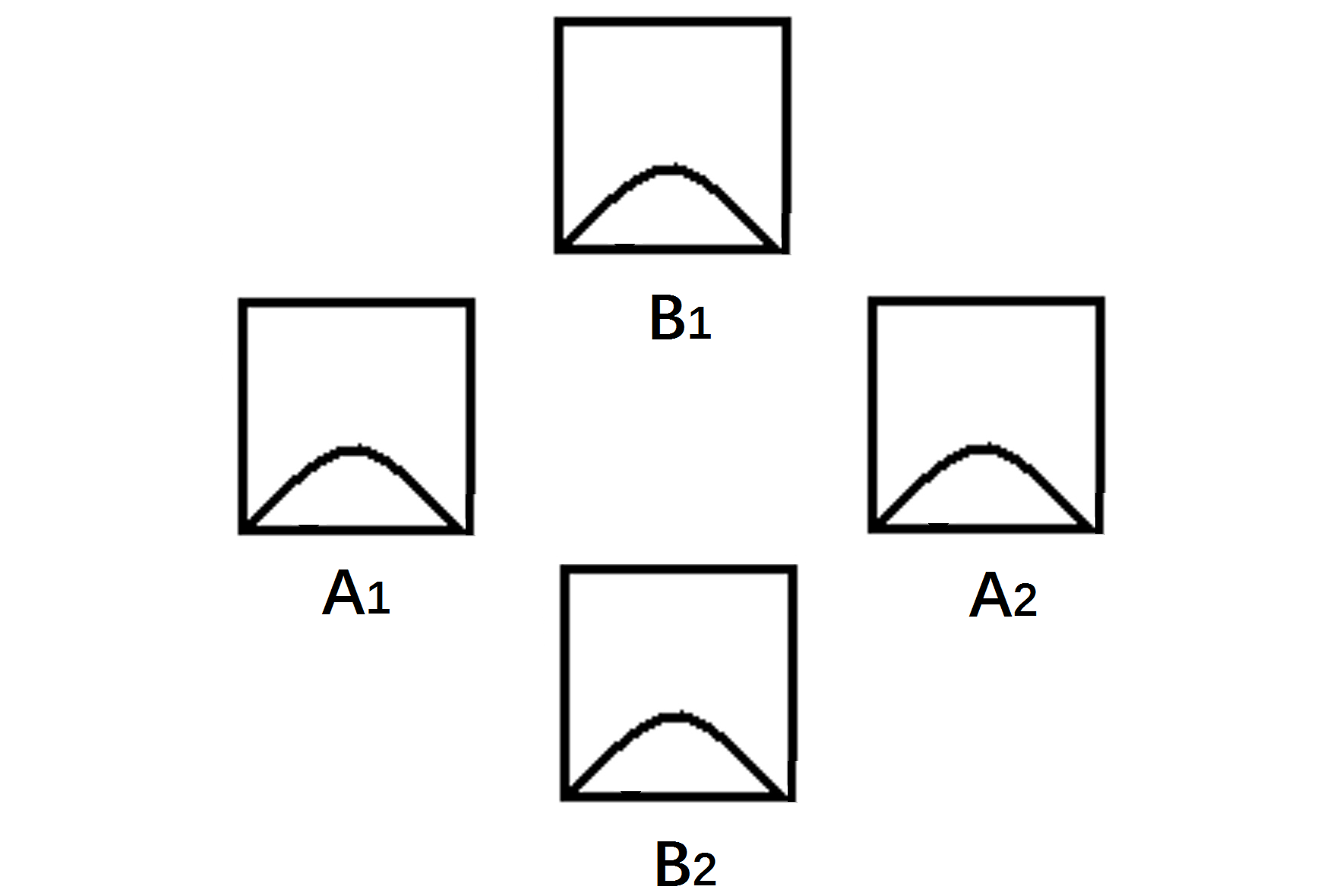}
\caption{Four separated fields in four identical boxes}
\end{center} 
\end{figure}

If wave functions represent physical fields, then at the initial instant there are four spatially separated fields: (1) the field of electron $A$ in box 1, denoted by $A_1$, which is represented by the wave function $\psi_1(x)$;  (2) the field of electron $A$ in box 2, denoted by $A_2$, which is represented by the wave function $\psi_2(x)$;  (3) the field of electron $B$ in box 3, denoted by $B_1$, which is represented by the wave function $\varphi_1(x)$; and (4) the field of electron $B$ in box 4, denoted by $B_2$, which is represented by the wave function $\varphi_2(x)$.\footnote{It has been argued that these fields have effective mass and charge distributions, which can be measured by protective measurements (Aharonov and Vaidman, 1993; Aharonov, Anandan and Vaidman, 1993; Gao, 2015, 2017).}  
According to the Schr\"{o}dinger equation, the fields of electron $A$, namely $A_1$ and $A_2$, will have (electromagnetic and gravitational) interactions with the fields of electron $B$, namely $B_1$ and $B_2$, but $A_1$ and $A_2$, as well as $B_1$ and $B_2$, have no interactions with each other, or the field of each electron has no self-interactions.\footnote{Note that the wave function of the two electrons will be still a product state during a protective interaction between them (Aharonov and Vaidman, 1993; Aharonov, Anandan and Vaidman, 1993; Gao, 2015, 2017).} 

This seems to pose a puzzle for the field ontologists. 
The puzzle has two aspects. 
First, if the fields of electrons $A$ and $B$ are in the same three-dimensional space  and the two electrons are identical in all aspects as usually thought, then it seems that the field of one electron cannot distinguish between itself and the field of the other electron. In this case, it is impossible that the field of one electron has interactions with the field of the other electron but has no self-interactions. 
This aspect of the puzzle may be called distinguishability of the fields of identical particles. 
Second, even though the field of one electron can distinguish between itself and the field of another electron, the distinguishability alone does not explain the non-existence of self-interactions. 
It is natural to expect that if the field of one electron has (electromagnetic and gravitational) interactions with the field of another electron, then  it will also have (electromagnetic and gravitational) self-interactions. 
This aspect of the puzzle may be called non-existence of self-interactions.
In the following, I will discuss three possible solutions of this puzzle a field ontologist  may provide. 

\section{Three solutions}

\subsection{Laws without explanations}

A natural defense strategy for the field ontologists seems to be appealing to the fundamentality of the Schr\"{o}dinger equation in quantum mechanics. 
Concretely speaking, the strategy is to appeal to the form of the Schr\"{o}dinger equation and say that it tells us that the field ontology has to behave in such a way without certain self-interactions, while the equation is what it is and no further explanation is needed.

Indeed, if we take the Schr\"{o}dinger equation as a fundamental law (in the case of the field interpretation of the wave function in the non-relativistic domain), then it will be not obvious that we need to explain the form of the equation. 
Fundamental laws of nature are supposed to be at the ground floor of explanations, and they are where explanations start. In other words, fundamental laws explain other facts and they themselves are not to be explained further. 

However, this solution seems unsatisfactory for two reasons. 
First, it can be argued that the puzzle cannot be dissolved by appealing only to the Schr\"{o}dinger equation. 
It is not that the Schr\"{o}dinger equation itself has some puzzling features, 
but that the field ontology has some puzzling features when the dynamics for the fields is described by the Schr\"{o}dinger equation, such as the distinguishability of the \emph{fields} of identical particles. 
The field ontology seems to imply that the fields of identical particles are indistinguishable. But the dynamics for the fields, namely the Schr\"{o}dinger equation says that these fields should be distinguishable, since according to the equation the field of one electron has interactions with the field of another electron but has no self-interactions. 
Thus, the field ontology needs to explain how the fields of identical particles can be distinguished; otherwise the proposal is at least incomplete. 

The second reason is a general consideration.  
Since the Schr\"{o}dinger equation is a dynamics for the ontology (when assuming a realist view), it is natural to expect that the form of the equation says something about what the ontology is. 
Different ontologies usually have different forms of dynamics after all. 
For example, Newtonian equations and Maxwell equations are different, and they are dynamics for two different ontologies, classical particles and classical fields, respectively.  
Thus, if the field ontology for the wave function has puzzling features that cannot be explained in a satisfactory way, then this might suggest that the ontology for the wave function is not fields. 
This makes a more careful analysis of these puzzling features necessary and even pressing. 


\subsection{Fields on a high-dimensional space}

Wave function realism provides a possible way to distinguish the fields of identical particles. 
According to this view, the wave function of $N$-body system is a real physical field on a fundamental $3N$-dimensional space (Albert, 1996, 2013, 2015). 
Then, there will be an ontological difference between the fields belonging to different electrons: the fields of different electrons (which are part of one universal field) live in different dimensions/subspaces of the $3N$-dimensional space. For the product state of the electrons $A$ and $B$, ${1 \over 2}[\psi_1(x_A)+\psi_2(x_A)][\varphi_1(x_B)+\varphi_2(x_B)]$, $x_A$ and $x_B$ are $3$-dimensional coordinates of a different subspace of the fundamental $3N$-dimensional space. 
Thus, the field of electron $A$ can distinguish between itself and the field of electron $B$ in principle by identifying which subspace it lives in. 

However, although wave function realism may explain the distinguishability of the fields of identical particles such as electrons, 
it does not provide an explanation of the non-existence of self-interactions. 
Even worse, it seems that this view further increases the difficulty to explain why the fields of different electrons (which live in different subspaces) have interactions, but the field of the same electron (which lives in the same subspace) has no self-interactions; it is arguable that living in different spaces will prevent, not facilitate, interactions. 

On the other hand, there have been plausible arguments supporting that the $3N$-dimensional space is actually an $N \times 3$-dimensional space, where the $3$-dimensional coordinates of different electrons are the same three dimensions (Monton, 2002, 2006, 2013; Lewis, 2004, 2013, 2016; Gao, 2017; Chen, 2017; Allori, 2018). 
In particular, when the two electrons $A$ and $B$ are in a product state in the above example, it seems obvious that their fields are in the boxes in the same three-dimensional space. 
If these arguments are indeed valid, then the fields of different particles will be in the same three-dimensional space. 
In this case, the distinguishability of the fields of identical particles is still in want of an explanation. 

To sum up, wave function realism seems to provide only limited resources to solve the above puzzle for the field ontologists. Even though it may explain the distinguishability of the fields of identical particles by assuming the existence of a fundamental high-dimensional space, it does not explain why the field of each particle has no (electromagnetic and gravitational) self-interactions. 

\subsection{Identical particles are not identical}

In the following, I will discuss how to solve the puzzle in three-dimensional space. 

A possible solution of the puzzle is to assume that 
the field of each electron has an identity marker belonging only to it, and moreover, the fields carrying different identity markers have interactions, while the fields carrying the same identity marker have no interactions.\footnote{I thank a referee of European Journal for Philosophy of Science for this insightful proposal.} 
Concretely speaking, each electron has an identity marker belonging only to it. 
The wave function corresponding to that electron (which the field ontologists take to be a real physical field) carries that identity marker with it wherever it goes. Furthermore, any potentials that the wave function generates (or whatever fields propagate from this field) also carry the same identity marker. 
Moreover, it is assumed that wave functions are affected by potentials bearing identity markers belonging to different electrons, while there is no effect when a wave function carrying a certain identity marker encounters a potential bearing the same identity marker. 

This proposal seems to provide a better solution of the puzzle of field ontology. 
However, it also has several potential problems. 
First, adding identity markers to electrons is inconsistent with our current understanding of the indistinguishability of identical particles. 
It is a basic postulate of modern physics that particles of the same type such as electrons are fundamentally indistinguishable from each other. 
For example, all electrons have identical physical properties: the same mass, same charge, same total spin, etc. Moreover, the indistinguishability of identical particles imposes a strong constraint on the form of the multi-particle wave functions: fermions such as electrons always have antisymmetric states, while bosons such as photons always have symmetric states. 
Although the existence of identity markers is not inconsistent with experience and these forms of the multi-particle wave functions (since these identity markers are unobservable and not described by the wave functions), 
it seems better not to violate the indistinguishability of identical particles at the fundamental level. 

Next, although adding identity markers to electrons can explain the first aspect of the above puzzle, namely the distinguishability of the fields of identical particles, it does not provide an explanation for the second aspect of the puzzle, namely the non-existence of self-interactions. 
Why do fields carrying the same identity marker have no interactions? 
No answer is provided by the above solution. 
In this sense, the solution, like wave function realism, is at least incomplete unless it also includes a further explanation for the absence of self-interactions. 

Finally, the above solution cannot explain the absence of interactions between different branches of an entangled state of many particles. 
Consider a spatially entangled state of the electrons $A$ and $B$: ${1\over {\sqrt{2}}}[\psi_1(x_A)\varphi_1(x_B)+\psi_2(x_A)\varphi_2(x_B)]$. 
For this state, there are still four fields in the four boxes: $A_1$ in box 1, $A_2$ in box 2, $B_1$ in box 3, and $B_2$ in box 4 (see Figure 1). 
According to the Schr\"{o}dinger equation, 
the fields $A_1$ and $B_1$ have (electromagnetic and gravitational) interactions, and the fields $A_2$ and $B_2$ have (electromagnetic and gravitational) interactions, but  the fields $A_1$ and $A_2$,  as well as the fields $A_1$ and $B_2$, have no interactions, and the fields $B_1$ and $B_2$,  as well as the fields $B_1$ and $A_2$, have no interactions. 
Thus, in this case, not only the fields carrying the same identity marker have no interactions, but also the fields carrying different identity markers may have no interactions either. This cannot be explained by the above solution. 

\section{The puzzle reconsidered}

The above analysis leads us to a more essential aspect of the puzzle,
namely that the fields in different (product state) branches of a spatially entangled state have no interactions, no matter these fields belong to the same particle or different particles. 
Obviously, adding identity markers to each particle cannot solve the puzzle. 
In the final analysis, the fields in different branches should be distinguishable at each point in space, so that they can be affected differently by the potentials at the point in different branches; when the potential is in the same branch as the field, the field is affected by the potential, while when the potential and the field are in different branches there is no effect. 

One might think that the puzzle may be solved in a similar way by adding identity markers to each branch of a spatially entangled superposition. 
But this solution seems to face more serious challenges.  
First, a spatially entangled superposition can be decomposed of infinitely many branches such as product states of position eigenstates of particles. 
Thus we need to add infinitely many identity markers to the theory.
Moreover, since a superposition can be decomposed with respect to infinitely many different bases, we actually need to add infinitely many different sets of identity markers to the theory. 
If these identity markers are not represented by known quantities in the theory, then this will make the revised theory very clumsy and unnatural. 
Next, when the branches of a superposition are created or annihilated during the time evolution of the superposition, we also need to add an additional dynamics to the theory, which is responsible for the assignment of identity markers. It is unclear what the dynamics is and whether it can assume a unique and simple form. 

Third, even though all these can be done, we still need to explain why the interactions between the fields in different branches of a spatially entangled state is screened or prevented and what the role of these identity markers is in this process. 
Note that the preventing mechanism applies to all interactions including both electromagnetic and gravitational interactions, although the nature of these interactions may be different. 
Only if all these problems can be solved in a satisfactory way, can we say that the puzzle has been solved for the field ontology. 


\section{A particle ontological way out}

Although we cannot exclude the possibility that the field ontologists may find a satisfactory way to solve the puzzle, the above analysis does suggest that we had better have a look at other possible ontologies. Maybe what the puzzle tries to tell us is that the ontology for the wave function is not fields. 
In the following, I will argue that the puzzle may be solved by a particle ontological interpretation of the wave function. 

As noted above, the puzzle is essentially composed of two parts: (1) the distinguishability of fields and potentials in different branches; and (2) the non-existence of influences of potentials on fields when they are in different branches. 
My suggested solution of this puzzle is to assume that 
when fields and potentials are in the same branch at a point in space, they exist there simultaneously, while when fields and potentials are in different branches at a point in space, they exist there at different instants in continuous time. 

Here is why this solution solves the puzzle. 
When fields and potentials in different branches exist at any point in space at different instants, they can be distinguished by the instants at which they exist at the point. 
This also means that the set of instants at which fields and potentials exist at a point in space in a branch may be regarded as their identity markers related to the branch.  
This explains the first part of the puzzle. 
Furthermore, since the fields and potentials in different branches do not exist simultaneously at any point in space, the fields are not affected by the potentials when they are in different branches. 
This explains the second part of the puzzle. 

It is worth noting that this solution also works when the potentials (which dissipate instantaneously from a field are replaced by the fields propagating with the speed of light from this field (which mediate the interactions), since it concerns only the influences of the potentials or the interaction mediating fields on the fields at each point in space. 
This means that the solution is valid in both non-relativistic and relativistic domains. 

Since each wave function can be decomposed of a superposition of position eigenstates, 
and according to the above solution, 
the supposed field is required to exist at different instants in any different branches, namely any different positions in this case, 
there is no continuous field spreading throughout space, but only a point-like entity or discrete particle being in a position at any instant, and its motion during an arbitrarily small time interval generates the effective field represented by the wave function of the particle.  

This will lead to a particle ontological interpretation of the wave function (Gao, 2017). Here the concept of particle is used in its usual sense. A particle is a small localized object with mass and charge, and it is only in one position in space at each instant. 
A few authors have suggested that the wave function represents a property of particles in three-dimensional space (see e.g. Monton, 2013; Lewis, 2013, 2016). But they do not give a concrete ontological picture of these particles in space and time and specify what property the property is. 
According to Gao's (2017) interpretation, an electron is a point-like particle, and the wave function of an electron is a description of the state of its motion, which is random and discontinuous in nature, and in particular, the modulus squared of the wave function gives the probability density that the electron appears in every possible position in space. At a deeper level, the wave function may be regarded as a description of the propensity property of the electron that determines its random discontinuous motion (RDM).  

\begin{figure}[h]\label{}
\begin{center} 
\includegraphics[scale=0.39]{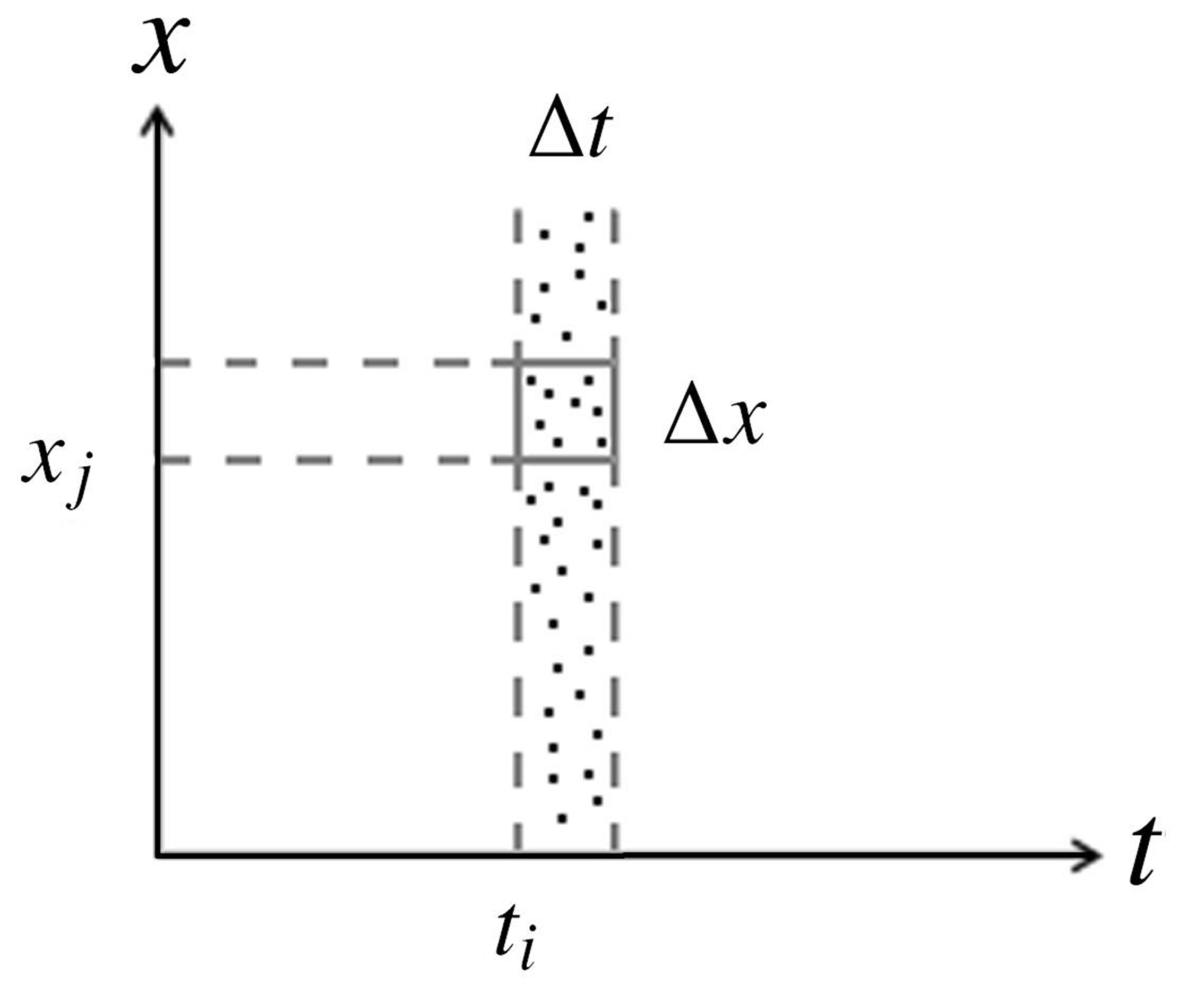}
\caption{A particle ontological interpretation of the wave function}
\end{center} 
\end{figure}

Here is a more detailed introduction of Gao's particle ontological interpretation of the wave function. 
Consider the state of one-dimensional RDM of a particle in finite intervals $\Delta t$ and $\Delta x$ around a space-time point ($t_i$,$x_j$) as shown in Figure 2. The positions of the particle form a random, discontinuous trajectory in this square region. 
We study the projection of this trajectory in the $t$-axis, which is a dense instant set in the time interval $\Delta t$. Let $W$ be the discontinuous trajectory of the particle and $Q$ be the square region $[x_j,x_j+\Delta x] \times [t_i,t_i+\Delta t] $. The dense instant set can be denoted by $\pi_t(W \cap Q) \in \Re$, where $\pi_t$ is the projection on the $t$-axis. According to the measure theory\index{measure theory}, we can define the Lebesgue measure:

\begin{equation}
M_{\Delta x,\Delta t}(x_j,t_i)=\int_{\pi_t(W \cap Q) \in \Re}dt.  
\label{}
\end{equation}

\noindent Since the sum of the measures of the dense instant sets in the time interval $\Delta t$ for all $x_j$ is equal to the length of the continuous time interval $\Delta t$, we have: 

\begin{equation}
\sum_j {M_{\Delta x,\Delta t}(x_j,t_i)}=\Delta t.
\label{m}
\end{equation}

\noindent Then we can define the measure density\index{set of points!measure density of} as follows:

\begin{equation}
\rho(x,t)=\lim_{\Delta x, \Delta t \rightarrow 0} {M_{\Delta x,\Delta t}(x,t)/(\Delta x \cdot \Delta t)}.
\label{rho}
\end{equation}

\noindent $\rho(x,t)$ may be called position measure density or position density in brief. 
This quantity provides a strict description of the position distribution of the particle in an infinitesimal space interval $dx$ around position $x$ during an infinitesimal interval $dt$ around instant $t$, and it satisfies the normalization relation $\int_{-\infty}^{+\infty}{\rho(x,t)dx}=1$ by (\ref{m}). 
Since the position density $\rho(x,t)$ changes with time in general, we may further define the position flux density $j(x,t)$ through the relation $j(x,t)=\rho(x,t)v(x,t)$, where $v(x,t)$ is the velocity of the local position density.\footnote{Here it may be worth emphasizing that $v(x,t)$ does not refer to the actual velocity of particles, which is infinite. This is in notable contrast to Bohm's theory, where $v(x,t)$ determines the actual velocity of particles moving along continuous trajectories.} It describes the change rate of the position density. Due to the conservation of measure, $\rho(x,t)$ and $j(x,t)$ satisfy the continuity equation: 

\begin{equation}
{{\partial \rho(x,t)}\over{\partial t}}+{{\partial j(x,t)}\over{\partial x}}=0.
\label{}
\end{equation}

\noindent The position density $\rho(x,t)$ and position flux density $j(x,t)$ provide a complete description of the state of RDM of a particle. 

This description of the motion of a particle can be extended to the motion of many particles. At each instant a quantum system of $N$ particles can be represented by a point in an $3N$-dimensional configuration space.
During an arbitrarily short time interval or an infinitesimal time interval around each instant, these particles perform random discontinuous motion in three-dimensional space, and correspondingly, this point performs random discontinuous motion in the configuration space. 
Then, similar to the single particle case, the state of the system can be described by the position density $\rho(x_1,x_2,...x_N,t)$ and position flux density $j(x_1,x_2,...x_N,t)$ defined in the configuration space\index{configuration space}. There is also the relation $\rho(x_1,x_2,...x_N,t)=\varrho(x_1,x_2,...x_N,t)$, where $\varrho(x_1,x_2,...x_N,t)$ is the probability density that particle 1 appears in position $x_1$ and particle 2 appears in position $x_2$ ... and particle $N$ appears in position $x_N$. When these $N$ particles are independent with each other, the position density can be reduced to the direct product of the position density for each particle, namely $\rho(x_1,x_2,...x_N,t)=\prod_{i=1}^N\rho(x_i,t)$.

Although the motion of particles is essentially discontinuous and random, the discontinuity and randomness of motion are absorbed into the state of motion, which is defined during an infinitesimal time interval around a given instant and described by the position density and position flux density. Therefore, the evolution of the state of RDM of particles may obey a deterministic continuous equation. 
By assuming the nonrelativistic equation of RDM is the Schr\"{o}dinger equation and considering the form of the resulting continuity equation,\footnote{It is also possible that the random motion of particles may affect the time evolution of the wave function and the Schr\"{o}dinger equation is replaced with a revised Schr\"{o}dinger equation with a stochastic evolution term describing the RDM as a stochastic process. This will  lead to collapse theories (see Gao, 2017 for more details).} we can obtain the relationship between the position density $\rho(x,t)$, position flux density $j(x,t)$ and the wave function $\psi(x,t)$. $\rho(x,t)$ and $j(x,t)$ can be expressed by $\psi(x,t)$ as follows:

\begin{equation}
\rho(x,t)=|\psi(x,t)|^2,
\label{}
\end{equation}

\begin{equation}
j(x,t)={\hbar \over{2mi}}[\psi^*(x,t){{\partial \psi(x,t)}\over{\partial x}}-\psi(x,t){{\partial \psi^*(x,t)}\over{\partial x}}].
\label{}
\end{equation}

\noindent Correspondingly, the wave function $\psi(x,t)$ can be uniquely expressed by $\rho(x,t)$ and $j(x,t)$ or $v(x,t)$ (except for an overall phase factor): 

\begin{equation}
\psi(x,t)=\sqrt{\rho(x,t)}e^{im\int_{-\infty}^{x}{v(x',t)}dx'/\hbar}.
\label{}
\end{equation}

\noindent In this way, the wave function $\psi(x,t)$ also provides a complete description of the state of RDM of a particle. 
A similar one-to-one relationship between the wave function and position density, position flux density also exists for RDM of many particles. For the motion of many particles, the position density and position flux density are defined in a $3N$-dimensional configuration space, and thus the many-particle wave function, which is composed of these two quantities, also lives on the $3N$-dimensional configuration space. 

It is well known that there are several ways to understand objective probability, such as frequentist, propensity, and best-system intepretations. In the case of RDM of particles, the propensity interpretation seems more appropriate. 
This means that the wave function in quantum mechanics should be regarded not simply as a description of the state of RDM of particles, but more suitably as a description of the instantaneous property of the particles that determines their RDM at a deeper level.
In particular, the modulus squared of the wave function represents the propensity property of the particles that determines the probability density that they appear in every possible group of positions in space.  
In contrast, the position density and position flux density, which are defined during an infinitesimal time interval around a given instant, are only a description of the state of the resulting RDM of particles, and they are determined by the wave function. In this sense, we may say that the motion of particles is ``guided" by their wave function in a probabilistic way. 
Note that this interpretation is more consistent with the theoretical formalism of quantum mechanics, in which the wave function (and potentials) are usually evaluated at a single instant. 

Now the question is: Can the RDM of particles explain the non-existence of self-interactions for a quantum system such as an electron? 
The answer is yes in the non-relativistic domain where quantum mechanics is valid. 
In the non-relativistic domain, since the (electromagnetic and gravitational) potentials of an electron dissipate from it instantaneously, the electron as a particle being in a position at an instant never encounters the potentials generated by it when it is in another position at another instant. 
In other words, the electron and the potentials it generates do not exist simultaneously at any point in space. 
For example, when electron $A$ moves between the two boxes in Figure 1 and is in one box at an instant, it does not encounter the potentials generated by it when it is in the other box at another instant. 
Then, similarly to solve the above puzzle, 
this explains why an electron has no self-interactions, e.g. why the two wavepackets of electron $A$ in the two boxes have no (electromagnetic and gravitational) interactions. 

However, it seems that there is an issue in the relativistic domain where quantum mechanics is replaced by quantum field theory. 
In the relativistic domain, the (electromagnetic and gravitational) fields of an electron propagate from it with the speed of light, and thus it seems possible that the electron as a particle undergoing RDM may encounter the fields generated by itself (see also Lazarovici, 2017). 
Then, why does this never happen so that an electron is affected by the fields generated by it?\footnote{Note that I omit the self-energy of an electron here, since after the divergent self-interaction is removed by renormalization, it is a small quantity compared with the supposed self-interactions I discuss in this paper.} 
In my view, we need to resort to the dynamics such as interacting quantum field theory, according to which such states are never formed. For example, the entangled superposition of an electron in two boxes and its fields is always a superposition in  which an electron in one box and the fields generated by it when it is in the other box are in different branches. Then, the RDM of particles can still provide an explanation of why the electron is not affected by its fields at each point in space: it is because they do not exist there simultaneously. 

\section{Conclusions}

It has been debated what the ontological content of quantum mechanics is. 
The field ontology is still a popular position among philosophers of physics. 
In this paper, I present a new analysis of the field ontology for the wave function. 
I argue that the non-existence of self-interactions for a quantum system such as an electron poses a puzzle for the field ontologists, and the three solutions they may provide are not fully satisfactory. Moreover, I also propose a solution of this puzzle in terms of particle ontology. 
Maybe it is time for us to take particles seriously in quantum mechanics.

\section*{Acknowledgments}
I am very grateful to the editors and referees of \emph{Foundations of Physics} for their insightful comments and helpful suggestions. This work
is supported by the National Social Science Foundation of China (Grant No.
16BZX021).

\section*{References}
\renewcommand{\theenumi}{\arabic{enumi}}
\renewcommand{\labelenumi}{[\theenumi]}
\begin{enumerate}

\item{} Aharonov, Y., Anandan, J. and Vaidman, L. (1993). Meaning of the wave function. Phys. Rev. A 47, 4616. 

\item{} Aharonov, Y. and Vaidman, L.   (1993). Measurement of the Schr\"{o}dinger wave of a single particle, Physics Letters A 178, 38.

\item{}  Albert, D. Z. (1996). Elementary Quantum Metaphysics. In J. Cushing, A. Fine and S. Goldstein (eds.), Bohmian Mechanics and Quantum Theory: An Appraisal. Dordrecht: Kluwer, 277-284.

\item{} Albert, D. Z. (2013). Wave function realism. In Ney and Albert (2013), pp. 52-57.

\item{} Albert, D. Z. (2015). After Physics. Cambridge, MA: Harvard University Press.

\item{} Allori, V. (2018). A New Argument for the Nomological Interpretation of the Wave Function: The Galilean Group and the Classical Limit of Nonrelativistic Quantum Mechanics. International Studies in the Philosophy of Science 31, 177-188.

\item{} Allori, V., S. Goldstein, R. Tumulka, and N. Zangh\`{i} (2008). On the common structure of Bohmian mechanics and the Ghirardi-Rimini-Weber theory, British Journal for the Philosophy of Science 59 (3), 353-389.

 \item  Allori, V., Goldstein, S., Tumulka, R., and Zanghi, N. (2014). Predictions and Primitive Ontology in Quantum Foundations: A Study of Examples. British Journal for the Philosophy of Science, 65, 323-52.

 \item Arntzenius, F. (2012). Space, Time, and Stuff. Oxford University Press.

 \item  Baker, D. J. (2016). The philosophy of quantum field theory. In Oxford Handbooks Online. Oxford University Press.

\item{} Bell, J. S. (1987).  Speakable and Unspeakable in Quantum Mechanics. Cambridge: Cambridge University Press.

\item{} Belot, G. (2012). Quantum states for primitive ontologists: A case study. European Journal for Philosophy of Science 2, 67-83.

 \item  Chen. E. K. (2017). Our fundamental physical space: An essay on the metaphysics of the wave function. The Journal of Philosophy, 114 (7), 333-365. 

\item{} Forrest, P. (1988). Quantum Metaphysics. Oxford: Blackwell.

\item Gao, S. (ed.) (2015). Protective Measurement and Quantum Reality: Toward a New Understanding of Quantum Mechanics. Cambridge: Cambridge University Press.

\item Gao, S. (2017). The Meaning of the Wave Function: In Search of the Ontology of Quantum Mechanics. Cambridge: Cambridge University Press.

\item{}  Ghirardi, G. C. (1997). Quantum dynamical reduction and reality: Replacing probability densities with densities in real space, Erkenntnis, 45, 349. 

\item{} Ghirardi, G. C. (2016). Collapse Theories, The Stanford Encyclopedia of Philosophy (Spring 2016 Edition), Edward N. Zalta (ed.), http://plato.stanford.edu/archives/spr2016/entries/qm-collapse/.

\item{} Ghirardi, G. C., Grassi, R. and Benatti, F. (1995). Describing the macroscopic world: Closing the circle within the dynamical reduction program. Foundations of Physics, 25, 313-328.

\item{} Goldstein, S. (2017). Bohmian Mechanics, The Stanford Encyclopedia of Philosophy (Summer 2017 Edition), Edward N. Zalta (ed.), https://plato.stanford.edu/archives/sum2017/entries/qm-bohm/.

 \item Hubert, M. and Romano, D. (2018). The wave-function as a multi-field.  European Journal for Philosophy of Science. 8(3), 521-37. 
 
 \item Ismael, J. and Schaffer, J. (2016). Quantum holism: nonseparability as common
ground. Synthese.	 https://link.springer.com/article/10.1007/s11229-016-1201-2.

 \item Lazarovici D. (2017). Review of Shan Gao's ``The Meaning of the Wave Function: In Search of the Ontology of Quantum Mechanics.'' International Studies in the Philosophy of Science 31(3), 321-24.

\item{} Lewis, P. J. (2004). Life in configuration space. British Journal for the Philosophy of Science 55, 713-729.

\item{} Lewis, P. J.  (2013). Dimension and illusion. In Ney and Albert (2013), pp. 110-125.

\item{} Lewis, P. J.  (2016). Quantum Ontology: A Guide to the Metaphysics of Quantum Mechanics. Oxford: Oxford University Press.

\item{} Maudlin, T. (2013). The nature of the quantum state. In Ney and Albert (2013), pp. 126-153.

\item{} Monton, B. (2002). Wave function ontology. Synthese 130, 265-277.
\item{} Monton, B. (2006). Quantum mechanics and 3N-dimensional space. Philosophy of Science 73, 778-789.
\item{} Monton, B. (2013). Against 3N-dimensional space. In Ney and Albert (2013), pp. 154-167.

\item Myvold, W. (2018). Ontology for Collapse Theories. In Gao, S. (ed.) Collapse of the Wave Function: Models, Ontology, Origin, and Implications. Cambridge: Cambridge University Press. pp. 97-123.

\item{} Ney, A. and D. Z. Albert (eds.) (2013). The Wave Function: Essays on the Metaphysics of Quantum Mechanics. Oxford: Oxford University Press.


\item{} Norsen, T. (2016). Quantum solipsism and non-locality. In M. Bell and S. Gao (eds.). Quantum Nonlocality and Reality: 50 Years of Bell’s theorem. Cambridge: Cambridge University Press.

 \item Swanson, N. (2018). How to be a Relativistic Spacetime State Realist. https://philsci-archive.pitt.edu/14412. Forthcoming in The British Journal for the Philosophy of Science. 

\item{} Wallace, D. (2012). The Emergent Multiverse: Quantum Theory according to the Everett Interpretation. Oxford: Oxford University Press.

\item{} Wallace, D.  and C. Timpson (2010). Quantum mechanics on spacetime I: spacetime state realism. British Journal for the Philosophy of Science 61, 697-727.

\end{enumerate}
\end{document}